\documentclass{JHEP3} 
\usepackage{epsfig}
\usepackage{subeqn}
\def\be{\begin{equation}}
\def\ee{\end{equation}}
\def\bc{\begin{center}}
\def\ec{\end{center}}
\def\bea{\begin{eqnarray}}
\def\eea{\end{eqnarray}}
\def\dd{\displaystyle}
\def\nn{\nonumber}
\def\ov{\overline}

\def\cL{{\cal L}}
\title{Equivalent effective Lagrangians
for Scherk-Schwarz compactifications}
\author{Carla Biggio 
\\ 
Dipartimento di Fisica `G.~Galilei', Universit\`a di Padova and
\\
INFN, Sezione di Padova, Via Marzolo~8, I-35131 Padua, Italy
\\ 
E-mail: \email{carla.biggio@pd.infn.it}}
\author{Ferruccio Feruglio \\
Dipartimento di Fisica `G.~Galilei', Universit\`a di Padova and
\\
INFN, Sezione di Padova, Via Marzolo~8, I-35131 Padua, Italy
\\ 
E-mail: \email{ferruccio.feruglio@pd.infn.it}}
\author{Andrea Wulzer \\
Dipartimento di Fisica, Universit\`a di Roma `La Sapienza' and
\\
INFN, Sezione di Roma, P.le Aldo Moro~2, I-00185 Rome, Italy
\\
E-mail: \email{andrea.wulzer@roma1.infn.it}}
\author{Fabio Zwirner \\
Dipartimento di Fisica, Universit\`a di Roma `La Sapienza' and
\\
INFN, Sezione di Roma, P.le Aldo Moro~2, I-00185 Rome, Italy
\\
E-mail: \email{fabio.zwirner@roma1.infn.it}}
\preprint{DFPD-02/TH/20 \\ ROME1-1339/02}
\abstract{
We discuss the general form of the mass terms that can appear in the
effective field theories of coordinate-dependent compactifications
{\em \`a la} Scherk-Schwarz. As an illustrative example, we consider
an interacting five-dimensional theory compactified on the orbifold
$S^1/Z_2$, with a fermion subject to twisted periodicity
conditions. We show how the same physics can be described by
equivalent effective Lagrangians for periodic fields, related by field
redefinitions and differing only in the form of the five-dimensional
mass terms. In a suitable limit, these mass terms can be localized at
the orbifold fixed points. We also show how to reconstruct the twist
parameter from any given mass terms of the allowed form. Finally,
after mentioning some possible generalizations of our results, we
re-discuss the example of brane-induced supersymmetry breaking in
five-dimensional Poincar\'e supergravity, and comment on its relation
with gaugino condensation in M-theory.
}
\begin{document} 
\section{Introduction and conclusions}
Traditionally, the effective field theories of coordinate-dependent
compactifications {\em \`a la} Scherk-Schwarz \cite{ss}, where the
periodicity conditions on the fields are twisted by a symmetry of the
action (or, more generally, of the equations of motion), were mostly
studied in the limit of generalized dimensional reduction. Only
recently the importance of keeping track of some crucial
higher-dimensional features of these compactifications was fully
appreciated. For example, in connection with the breaking of local
symmetries, an important feature of the Scherk-Schwarz mechanism is
the non-locality of the order parameter \cite{nonloc} (for a recent,
pedagogical discussion see also \cite{msss}), which improves the
ultraviolet behavior of symmetry-breaking quantities
\cite{uvsoft}. Such behavior would be out of theoretical control if
one were to consider the reduced four-dimensional effective theory for
the light modes only, instead of the compactified higher-dimensional
one with its full tower of Kaluza-Klein excitations.

Recently, a generalized formulation of the Scherk-Schwarz mechanism
was discovered \cite{bfz1}, in the context of five-dimensional (5D)
field-theory orbifolds, which can generate localized mass terms for
the higher-dimensional bulk fields at the orbifold fixed
points. Noticeable applications were found to `brane-induced'
supersymmetry breaking \cite{bfz2} (see also \cite{mq}) and to gauge
symmetry breaking \cite{bf}. The resulting mass spectrum is
characterized by a universal shift of the Kaluza-Klein levels, exactly
as in the conventional formulation. Also, when applied to minimal 5D
Poincar\'e supergravity, both the conventional and the new formulation
lead, at the classical level, to vanishing vacuum energy, undetermined
gravitino masses (with a flat direction associated to the
compactification radius $R$), and goldstinos along the internal
components of the 5D gravitinos.  Moreover, in both cases the mass
shifts are controlled by a global parameter, which for brane-induced
supersymmetry breaking can be identified with the overall jump of the
gravitino field across the orbifold fixed points. All these analogies
suggest that the two formulations may represent two equivalent
descriptions of the same physical system, and that this equivalence
may survive in the presence of interactions. This intepretation may
sound counterintuitive. It is often assumed that the Scherk-Schwarz
mechanism must correspond to a $y$-independent mass term in a basis of
periodic 5D fields.  It may also be tempting to associate the
`profiles' of $y$-dependent mass terms, directly linked to the shapes
of the mass eigenmodes, to some intrinsic physical property of the
system, and to expect that, when interactions are accounted for,
different shapes would unavoidably lead to different transition rates
and probabilities. We then believe that a systematic discussion of the
structure of the interacting 5D effective theories, originated by
Scherk-Schwarz compactifications on field-theory orbifolds, is in
order. This is the purpose of the present paper. In full agreement
with \cite{bfz1,bfz2,mq,bf}, we confirm the possibility of different
but equivalent forms for the mass terms in otherwise identical
effective theories, as a consequence of the freedom of performing
local field redefinitions without changing the physics. We classify
the conditions to be satisfied by the 5D mass terms, in a basis of
periodic fields, in order to be fully ascribed to a Scherk-Schwarz
twist, and we discuss how the equivalence continues to hold in the
presence of interactions.

The plan of the paper is the following. For simplicity, and for an
easier contact with the existing literature, we initially focus on the
theory of a massless 5D spinor, compactified on the orbifold $S^1/Z_2$
with twisted periodicity conditions. After recalling the general
formalism \cite{ss} and some well-known consistency conditions
\cite{const,conft,aqgau} for such a construction, we discuss the
allowed structures for the coordinate-dependent mass terms in the
corresponding effective 5D theories, formulated in a general basis of
canonically normalized periodic fields. In particular, we show that
the standard formalism, in which the Scherk-Schwarz twist is converted
into a constant and $Z_2$-even 5D mass term, corresponds to the choice
of a special basis of periodic fields within an infinite
class. Conversely, we show that, given an effective 5D theory with
periodic fields and an allowed set of coordinate-dependent mass terms,
we can move to an equivalent 5D theory with no mass terms and a
suitable Scherk-Schwarz twist. We conclude by mentioning how the
equivalence can be generalized to a very large class of theories with
arbitrary field content and interactions. As an example, we re-discuss
the application to brane-induced supersymmetry breaking \cite{bfz2,mq}
in 5D Poincar\'e supergravity \cite{pure5d}, and comment on its
relation with gaugino condensation \cite{hor,bulk,mno1,mno} in
M-theory \cite{hw}. Some useful definitions and formulae on
path-ordered integrals are collected in the Appendix.

\section{Twisted periodicity conditions for a 5D massless fermion on
$S^1/Z_2$}
We consider a 5D field theory compactified on the orbifold
$S^1/Z_2$. We gauge-fix the invariance under 5D general coordinate
transformations and choose the space-time coordinates $x^M \equiv
(x^m, y)$ in such a way that the background metric is the Minkowski
one, $\eta_{MN} = diag (-1, +1, +1, +1, +1)$. Similarly, we gauge-fix
the local 5D Lorentz invariance to put the background value of the
f\"unfbein in standard form, $e_M^{\; A} = \delta_M^A$. We can
represent the orbifold on the whole real axis, identifying points
related by a translation $T$ and a reflection $Z_2$ about the origin:
\be 
T: \; y \to y + 2 \pi R \, ,
\;\;\;\;\;\;\;\;\;\;
Z_2 : \;  y \to -y \, ,
\label{tz2}
\ee
where $R$ is the radius of $S^1$. 

For definiteness, we focus here on the theory of a 5D massless spinor
$\Psi(x^m,y)$, and ignore the gravitational degrees of freedom
associated with the fluctuations of the 5D metric. As will be
discussed in the final section, our results can be easily extended to
more general situations, including the case of the gravitino field in
5D supergravity, relevant for the discussion of local supersymmetry
breaking.  In terms of representations of the four-dimensional (4D)
Poincar\'e group, the residual invariance of the chosen space-time
background, the 5D spinor~\footnote{From now on, the $x^m$-dependence
of the fields will be always understood, and their $y$-dependence
indicated only when appropriate. Notice that we differ from other
frequently used notations in which $\Psi$ is represented by a Dirac
spinor ($\psi_2 \to \ov{\psi_2}$) or by a symplectic Majorana spinor.
We also define $\ov{\Psi} \equiv (\ov{\psi_1} \; \ov{\psi_2})^T$. Our
4D conventions are the same as in \cite{WB}.} $\Psi$ consists of two
Weyl spinors $\psi_{i}$ $(i=1,2)$:
\be 
\Psi = \left(
\begin{array}{c}
\psi_{1} \\
\psi_{2}
\end{array}
\right) \, .
\label{notation}
\ee

The system under consideration is described by the 5D Lagrangian:
\be
\cL=\cL_0 + \cL_{int} \, .
\label{lag}
\ee
In Eq.~(\ref{lag}), $\cL_0$ represents the free massless Lagrangian
for $\Psi$, 
\be
\cL_0=
i \ov{\Psi}^T \ov{\sigma}^m \partial_m \Psi -
\frac{1}{2} \left( i \, \Psi^T \widehat{\sigma}^2 
\partial_y \Psi + {\rm h.c.}\right) \, ,
\label{lag0}
\ee
where here and in the following we will denote with a hat the Pauli
matrices acting on objects such as the one in Eq.~(\ref{notation}).
The remaining part of the 5D Lagrangian of Eq.~(\ref{lag}),
$\cL_{int}$, contains possible interaction terms for the field $\Psi$,
and in general may depend on additional 5D fields.

With respect to the $Z_2$ reflection that defines the orbifold,
we will adopt the parity assignment
\be
\Psi(-y) = Z~ \Psi (y) \, ,
\;\;\;\;\;\;\;\;\;
Z= \widehat{\sigma}^3 \, .
\label{parity}
\ee
For the consistency of the orbifold construction, both $\cL_0$
and $\cL_{int}$ must be invariant under $Z_2$, when suitable  
parities are assigned to the fields other than $\Psi$ appearing
in $\cL_{int}$. The free Lagrangian $\cL_0$ is also invariant under 
\be
\Psi'(y) = U \, \Psi(y) \, ,
\label{su2}
\ee
where $U$ is a global $SU(2)$ transformation. We require that also
$\cL_{int}$ is $SU(2)$ invariant (after assigning suitable $SU(2)$
transformation properties to the fields other than $\Psi$) and does not
contain derivatives of the field $\Psi$ or of other fields with
non-trivial $SU(2)$ transformation properties.

In this framework, the field $\Psi(y)$ does not need to be periodic in
$y$. It can be periodic up to a global SU(2) transformation, with
`twisted' periodicity conditions \cite{ss}:
\be
\Psi(y+2 \pi R)=U_{\vec{\beta}} \Psi(y) \, ,
\;\;\;\;\;\;\;\;\;
U_{\vec{\beta}} \equiv 
\displaystyle{ e^{i \vec{\beta} \cdot \vec{\sigma} } }
= \cos \beta \, {\mathbf 1} + {\sin \beta \over \beta}
\, \vec{\beta} \cdot \vec{\sigma} \, ,
\label{twisted}
\ee
where $\vec{\sigma} = (\widehat{\sigma}^1, \widehat{\sigma}^2,
\widehat{\sigma}^3)$, $\vec{\beta} = (\beta^1, \beta^2, \beta^3)$ is a
triplet of real parameters, $\beta \equiv \sqrt{\beta_1^2 + \beta_2^2
+ \beta_3^2}$, and it is not restrictive to assume $\beta \le
\pi$. The operators $Z$ and $U_{\vec{\beta}}$ acting on the fields
should provide representations of the space-time operations $T$ and
$Z_2$ of Eq.~(\ref{tz2}). This gives rise to the well-known
consistency condition \cite{const,conft,aqgau}
\be
U_{\vec{\beta}} \, Z \, U_{\vec{\beta}} = Z \, ,
\label{cc}
\ee
and for $Z=\widehat{\sigma}^3$ implies~\footnote{The choice
$\vec{\beta}=(0,0,\pi)$ gives $U_{\vec{\beta}} = - {\mathbf 1}$, as
any other choice with $\beta = \pi$.}
\be
\vec{\beta}=(\beta_1,\beta_2,0) \, .
\label{beta}
\ee

If all fields in $\cL_{int}$ have trivial background values as
solutions of the corresponding classical equations of motion, then the
4D modes have a classical spectrum characterized by a universal shift
of the Kaluza-Klein levels, with respect to the mass eigenvalues $n/R$
of the periodic case, controlled by $\beta$:
\be
m= \frac{n}{R} - \frac{\beta}{2\pi R} \, ,
\;\;\;\;\;\;\;\;\;\;
(n \in {\mathbf Z}) \, .
\label{masses}
\ee
The corresponding eigenfunctions are, in the notation of
Eq.~(\ref{notation}) and with the periodicity conditions of
Eq.~(\ref{twisted}):
\be 
\Psi^{(n)}(y)= \chi(x) \, e^{\dd{\, i \, \gamma  \, \widehat{\sigma}^3}}  
\left(
\begin{array}{c}
\cos m y\\
\sin m y
\end{array}
\right) \, ,
\label{eigfun}
\ee
where $\chi(x)$ is $y$-independent 4D Weyl spinor satisfying the
equation $i \sigma^m \partial_m \, \ov{\chi} = m \, \chi$, and,
barring the trivial case $\vec{\beta} = 0$ in which the rotation 
angle $\gamma$ is arbitrary:
\be
\gamma = 
{1 \over 2} \arctan \left( \beta_1 \over \beta_2 \right) 
+ \delta + \rho \pi \,,
\;\;\;\;\;
\delta = \left\{
\begin{array}{lcc}
0      & {\rm for} & \beta_2 \ge 0 \\
\pi/2  & {\rm for} & \beta_2<0 
\end{array}
\right. \, ,
\;\;\;\;\;
(\rho \in {\mathbf Z}) \, .
\label{gamma}
\ee
All this is well-known, it was reported here only for completeness
and to make the following discussion more transparent.

\section{Generation of 5D mass terms for periodic fields}
We now move to a class of equivalent descriptions of the system
characterized so far by the Lagrangian of Eqs.~(\ref{lag}) and
(\ref{lag0}) and by the twisted periodicity conditions of
Eqs.~(\ref{twisted})-(\ref{beta}). Exploiting the fact that S-matrix
elements do not change if we perform a local and non-singular field
redefinition (see, e.g., \cite{redfs}), we replace the twisted fields
$\Psi(y)$ by periodic ones $\widetilde{\Psi}(y)$:
\be
\Psi(y)  =  V(y) \, \widetilde{\Psi}(y) \, ,
\;\;\;\;\;\;\;\;
\widetilde{\Psi} (y+2\pi R)  =  
\widetilde{\Psi} (y) \, ,
\label{redef}
\ee
where $V(y)$ must then be a $2\times 2$ matrix satisfying
\begin{subequations}
\label{reqall}
\be
V(y+2 \pi R)=U_{\vec{\beta}} \, V(y) \, ,
\label{req1}
\ee
\end{subequations}
as can be immediately checked from Eqs.~(\ref{twisted}) and
(\ref{redef}). Besides condition (\ref{req1}), we will impose for our
convenience two additional constraints on the matrix $V(y)$. One is
\addtocounter{equation}{-1}
\begin{subequations}
\addtocounter{equation}{1}
\be
V(y)\in SU(2) \, ,
\label{req2}
\ee
\end{subequations}
which guarantees that the redefinition is non-singular, and that the
kinetic terms for $\widetilde{\Psi}(y)$ remain canonical, as in
$\cL_0$.  Moreover, the interaction terms in $\cL_{int}$ are not
modified, even if the $SU(2)$ transformation is $y$-dependent, as long
as they do not involve derivatives of $\Psi$ or of other fields with
non-trivial $SU(2)$ transformation properties. Thus, new terms can
only originate from $\cL_0$, when the $y$-derivative acts on $V(y)$.
We also require that the new fields $\widetilde{\psi}_1(y)$ and
$\widetilde{\psi}_2(y)$ have the same parities as the original ones
${\psi}_1 (y)$ and ${\psi}_2 (y)$:
\addtocounter{equation}{-1}
\begin{subequations}
\addtocounter{equation}{2}
\be
\left\{
\begin{array}{rclcc}
V_{ij}(-y)&=&+V_{ij}(y)  & \phantom{bla} &(ij=11,22)\\
V_{ij}(-y)&=&-V_{ij}(y)  & \phantom{bla} &(ij=12,21)
\end{array} \right. \, .
\label{req3}
\ee
\end{subequations}
Notice that Eq.~(\ref{req3}) implies $V(0)=\exp \, ( \, i \, \theta \,
\widehat{\sigma}^3 \, )$, with $\theta \in {\mathbf R}$.

Before exploring the effects of the field redefinition of
Eq.~(\ref{redef}), we observe that the solution to the conditions
(\ref{reqall}) is by no means unique. Starting from any given solution
$V(y)$, a new set of solutions $V'(y)$ can be generated via matrix
multiplication:
\be
V'(y) = W_L(y) \, V(y) \, W_R(y) \, ,
\label{newsol}
\ee
provided that the following conditions are satisfied:
\begin{subequations}
\be
W_L(y+2\pi R) \, U_{\vec{\beta}} = U_{\vec{\beta}} \, W_L(y)
\, , \;\;\;\;\; W_R(y+2\pi R) = W_R(y) \, ,
\label{wy1}
\ee
\be
W_{L,R}(y)\in SU(2) \, ,
\label{wy2}
\ee
\be
\left\{
\begin{array}{rclcc}
(W_{L,R})_{ij}(-y)&=&+(W_{L,R})_{ij}(y)& \phantom{bla} &(ij=11,22)\\
(W_{L,R})_{ij}(-y)&=&-(W_{L,R})_{ij}(y)& \phantom{bla} &(ij=12,21)
\end{array} \right. \, .
\label{wy3}
\ee
\end{subequations}

We are now ready to explore the effects of the field redefinition of
Eq.~(\ref{redef}). The Lagrangian $\cL$, expressed in terms of the
periodic field $\widetilde{\Psi}(y)$, describes exactly the same
physics as before, but its form is now different:
\bea 
\cL(\Psi,\partial\Psi) & = &
\cL(\widetilde{\Psi},\partial\widetilde{\Psi}) + \left\{
-\frac{i}{2}\left[m_1(y)+i \, m_2(y)\right] \widetilde{\psi}_1
\widetilde{\psi}_1 \right. \nn \\ & + & \left.
\frac{i}{2}\left[m_1(y)-i \, m_2(y)\right]
\widetilde{\psi}_2 \widetilde{\psi}_2 + i \, m_3(y) \,
\widetilde{\psi}_1 \widetilde{\psi}_2 + {\rm h.c.} \right\} \, ,
\label{result}
\eea
where the mass terms $m_a(y)$ $(a=1,2,3)$ are the coefficients
of the Maurer-Cartan form
\be
m(y) \equiv m_a(y) \, \widehat{\sigma}^a = 
- i \, V^\dagger(y) \partial_y V(y) \, ,
\label{mcform}
\ee
and satisfy:
\begin{subequations}
\label{mmmm}
\be
m_a(y+2\pi R) = m_a(y) \, , 
\label{I}
\ee
\be
m_a(y) \in {\mathbf R} \, ,  
\label{II}
\ee
\be
m_{1,2}(-y) = + m_{1,2}(y) \, , 
\;\;\;\;\;
m_{3}(-y) = -m_{3}(y) \, ,
\label{III}
\ee
$$
U_{\vec{\beta}}  =   V(0) \,
P \left[ \exp \left( {\dd i \, 
\int_0^y dy' m(y')}
\right) \right] \, 
P_< \left[ \exp \left( {\dd i \, 
\int_y^{y+2\pi R} dy' m(y')}
\right) \right] \, \times \phantom{blablabla} 
$$
\be
P\,' \left[
\exp \left( {\dd - i \, \int_0^y 
dy' m(y')}\right)\right] 
V^\dagger(0) \, ,
\;\;\;\;\;
\left\{
\begin{array}{cccc}
P=P_< \, , &  P\,'=P_> & {\rm for} & y>0  \\
P=P_> \, , &  P\,'=P_< & {\rm for} & y<0 
\end{array}
\right. \; .
\label{IV}
\ee
\end{subequations}

Properties (\ref{I})-(\ref{III}) are in one-to-one correspondence with
conditions (\ref{req1})-(\ref{req3}) on $V(y)$. Eq.~(\ref{IV}) is
related with Eq.~(\ref{req1}), and prescribes how the information on
the twist of the original fields $\Psi(y)$ is encoded in the new
Lagrangian. The symbols $P_<$ and $P_>$ denote inequivalent
definitions of path-ordering, specified in the Appendix with some
useful properties and the proof of Eq.~(\ref{IV}). Notice that, by
taking the trace of both members in Eq.~(\ref{IV}), we obtain a
relation between the Wilson loop and the twist parameter $\beta$:
\be
\cos \beta = 
{1 \over 2} {\rm tr} \, U_{\vec{\beta}} = 
{1 \over 2}{\rm tr} \, P_< \left[ \exp \left( {\dd i \, 
\int_y^{y+2\pi R} dy' m(y')} \right) \right] \, .
\label{wloop}
\ee
Notice also that, because of the freedom of performing global $SU(2)$
transformations with the constant matrix $V(0)$, which are invariances
of the Lagrangian, different values of the twist $\vec{\beta}$ with
the same value of $\beta$ correspond to physically equivalent
descriptions.

The mass terms in Eq.~(\ref{result}) are of three different types,
associated with the three possible bilinears~\footnote{At variance
with ref.~\cite{mno}, we find that the coefficients of the bilinears
$\widetilde{\psi}_1 \widetilde{\psi}_1$ and $\widetilde{\psi}_2
\widetilde{\psi}_2$, although related, should not be necessarily
equal. We also disagree with the statement in \cite{mno} that `the
Scherk-Schwarz mechanism is equivalent to the mechanism of adding a
mass term only if this mass term is $Z_2$-even and constant'.}
$\widetilde{\psi}_1 \widetilde{\psi}_1$, $\widetilde{\psi}_2
\widetilde{\psi}_2$ and $\widetilde{\psi}_1
\widetilde{\psi}_2$. Because of Eq.~(\ref{mmmm}), they do not
correspond to the most general set of $y$-dependent mass terms allowed
by 4D Lorentz invariance, which would be characterized by three
independent complex functions. The r\^ole of the conditions
(\ref{mmmm}) is to guarantee the equivalence between the descriptions
on the two sides of Eq.~(\ref{result}). We stress again that this
equivalence is not limited to the free-field case, but also holds true
in the interacting case~\footnote{Actually, the equivalence between
two Lagrangians related by a local field redefinition holds
irrespectively of the explicit form assumed by the interaction
terms. From this point of view, we could drop the assumption that
$\cL$ does not contain derivatives acting on fields with non-trivial
$SU(2)$ transformation properties such as $\Psi$. In this case,
interaction terms involving $\partial_y \Psi$ would generate, via the
redefinition of Eq.~(\ref{redef}), additional but controllable
contributions to the right hand side of Eq.~(\ref{result}).}.

Also the converse is true. Given a Lagrangian such as the one on the
right-hand side of Eq.~(\ref{result}), expressed in terms of periodic
fields $\widetilde{\psi}_i(y)$ $(i=1,2)$ and with mass terms
satisfying Eq.~(\ref{mmmm}), we can move to the equivalent Lagrangian
of Eqs.~(\ref{lag}) and (\ref{lag0}), where all mass terms have been
removed, and the fields satisfy the generalized periodicity conditions
of Eq.~(\ref{twisted}), by performing the field redefinition of
Eq.~(\ref{redef}). As shown in the Appendix, $V(y)$ is given by:
\be
V(y) = V(0) \, P \left[ \exp \left( {\dd i 
\, \int_0^y dy' m(y') } \right) \right] \, ,
\;\;\;\;\;
P = \left\{
\begin{array}{ccc}
P_< & {\rm for} & y > 0 \\
P_> & {\rm for} & y < 0
\end{array}
\right.
\, .
\label{converse}
\ee
For any $V(0) = \exp \, ( \, i \, \theta \, \widehat{\sigma}^3 \, )$
($\theta \in {\mathbf R}$), conditions (\ref{reqall}) are satisfied
with $U_{\vec{\beta}}$ given by Eq.~(\ref{IV}).  The arbitrariness in
$V(0)$ reflects the fact that physically distinct theories are
characterized by $\beta$, not by $\vec{\beta}$.

\section{Examples and localization of 5D mass terms}
{}From the discussion of the previous section, it is clear that mass
`profiles' $m_a(y)$ for periodic fields, of the type specified in
Eq.~(\ref{mmmm}), do not have an absolute physical meaning. They can
be eliminated from the Lagrangian and replaced by a twist, the two
descriptions being completely equivalent. Moreover, all Lagrangians
with the same $\cL_{int}$ and mass profiles corresponding to the same
twist $\vec{\beta}$, as computed from Eq.~(\ref{IV}), are just
different equivalent descriptions of the same physics. Indeed, suppose
that $\cL^1$ and $\cL^2$ are two such Lagrangians, and call $V^I(y)$
$(I=1,2)$ the local redefinitions mapping $\cL^I$ into the Lagrangian
$\cL$ for the twisted, massless 5D fields $\Psi(y)$. Then $\cL^1$ and
$\cL^2$ are related by the local non-singular field redefinition $V(y)
= {V^2}^\dagger(y) V^1(y)$. This shows that, in the class of
interacting models under consideration, what matters is the twist
$\vec{\beta}$ and not the specific form of the mass terms $m_a(y)$
enjoying the properties (\ref{mmmm})~\footnote{Actually, in view of
the observations after Eqs.~(\ref{wloop}) and (\ref{converse}), what
really matters is $\beta$.}.  We can make use of this freedom to show
that $m_{1,2}(y)$ can be localized at the fixed points $y=0$ and/or
$y=\pi R$, without affecting the physical properties of the theory.

As an example, we consider the simple case in which the twist parameter
is just 
\be
\vec{\beta}= (0,\beta,0) \, .
\label{twist2}
\ee 
Then a frequently used solution to Eq.~(\ref{reqall}), for
the twist specified by Eq.~(\ref{twist2}), is
\be
V^O(y) = \exp \left( i \, \beta \widehat{\sigma}^2 
\frac{y}{2\pi R} \right) \, ,
\label{vss}
\ee
the symbol `$O$' standing for {\em `ordinary'}. Starting from the
Lagrangian $\cL^0$ for the periodic fields $\widetilde{\Psi}(y)$,
defined by $V^O(y)$ via the redefinition of Eq.~(\ref{redef}), and
performing the standard Fourier decomposition of the 5D fields into 4D
modes, we can immediately check that the 4D mass eigenvalues and
eigenfunctions are indeed given by Eqs.(\ref{masses})-(\ref{gamma}),
with $\gamma=\delta=0$. Applying Eq.~(\ref{mcform}) to $V^O(y)$, we
find the constant mass profile:
\be
m_1^O(y)=m_3^O(y)=0 \, ,
\;\;\;\;\;
m_2^O(y)=\frac{\beta}{2 \pi R} \, ,
\label{mo}
\ee
and we can check that, in agreement with Eq.~(\ref{wloop}):
\be
\beta = \int_y^{y+2\pi R}d y' m_2^O(y') \, .
\label{sametwist}
\ee

We now move, following \cite{bfz1}, to a more general solution of
Eqs.~(\ref{reqall}) and (\ref{twist2}), where, in a basis of periodic
fields, the system is described by a different Lagrangian $\cL^G$ (the
symbol `$G$' stands for {\em `generalized'}). $\cL^G$ is still of the
general form of Eq.~(\ref{result}), including interaction terms, but
now:
\be
m_1^G(y)=m_3^G(y)=0 \, , 
\;\;\;\;\;
m_2^G(y)\ne 0 \, ,
\label{mgen}
\ee
and $m_2^G(y)$ is an otherwise arbitrary real, periodic, even function
of $y$, with the property that
\be
\int_y^{y+2\pi R}d y' m_2^G(y') = 
\int_y^{y+2\pi R}d y' m_2^O(y') = 
\beta \, .
\ee
As long as the above properties are satisfied, the two Lagrangians
$\cL^O$ and $\cL^G$ are physically equivalent. Two representative and
equivalent choices of $m_2^G(y)$ are illustrated in Fig.~\ref{m2y}:
\FIGURE{
\epsfig{figure=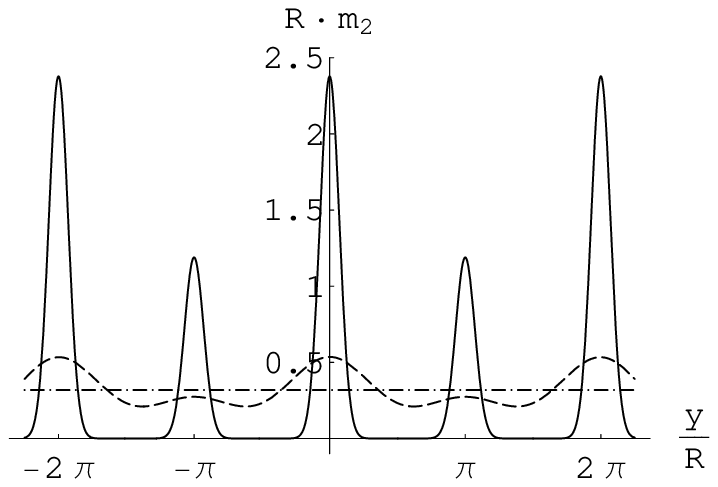,height=7.5cm}
\caption{Two representative and equivalent choices for $m_2^G(y)$,
corresponding to $\beta=2$. For reference, the dash-dotted line
shows the equivalent constant profile $m_2^O(y)= 1 / (\pi R)$.}
\label{m2y}
}
the dashed line shows a mild (Gaussian) localization around the
orbifold fixed points, the solid line a strong localization. The
interactions between $\widetilde{\Psi}(y)$ and other fields are not
determined by the shapes of the fermion eigenmodes and, indirectly, by
the profile of $m_2(y)$. Neither the mass spectrum, nor the
interactions depend on shapes, which are an artifact of the choice of
field variables. As long as the twist is kept fixed, shapes can be
arbitrarily deformed along $y$, without changing the physics.

A possible special choice for $m_2^G(y)$ is the singular limit:
\be
m_2^G(y)=\sum_{q=-\infty}^{+\infty}
\left[\delta_0 \, \delta(y-2 q \pi R)+
\delta_\pi \, \delta(y-(2 q+1) \pi R)\right] \, ,
\;\;\;\;\;
\delta_0+\delta_\pi=\beta \, ,
\label{mg}
\ee
where what we actually mean is, as discussed in detail in
\cite{bfz1,bfz2}, a suitably regularized version of the distribution
in Eq.~(\ref{mg}). This description is apparently quite remote from
the `ordinary' one. The mass terms vanish everywhere but at the
orbifold fixed points, where there are localized contributions to
$m_2(y)$. The redefinitions bringing from the massive periodic fields
of $\cL^O$ and $\cL^G$ to the corresponding massless twisted 5D fields
are:
\be
\Psi(y)=V^{O,G}(y)~ \widetilde{\Psi}^{O,G}(y) \, ,
\label{redefog}
\ee
with $V^O(y)$ given by Eq.~(\ref{vss}) and
\be
V^G(y)=\exp \left[i \, \alpha(y) \widehat{\sigma}^2\right] \, .
\label{redefg}
\ee
Here $\alpha(y)$, depicted in Fig.~\ref{aofy} for some representative
choices of $\delta_0$ and $\delta_\pi$, is given by:
\FIGURE{
\epsfig{figure=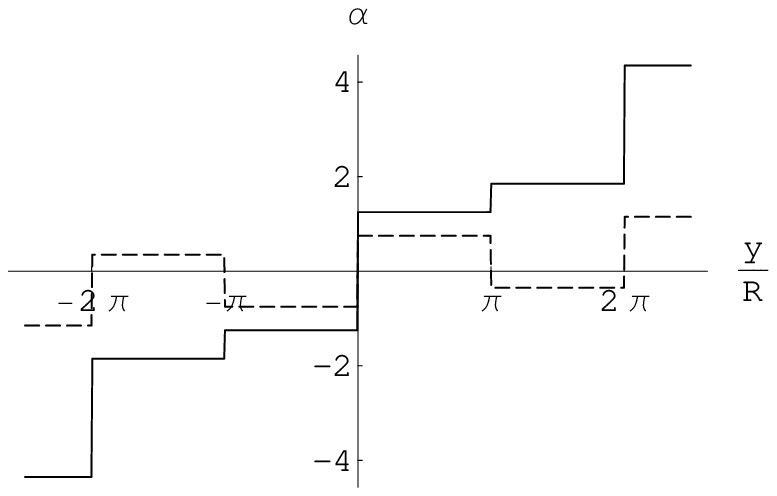,height=7.5cm}
\caption{The function $\alpha(y)$ for two representative parameter
choices: the solid line corresponds to $\delta_0 = 2.5$,
$\delta_\pi = 0.6$, the dashed one to $\delta_0 = 1.5$,
$\delta_\pi = -1.1$.}
\label{aofy}
}
\be
\alpha(y)=\frac{\delta_0-\delta_\pi}{4}~\epsilon(y)
+\frac{\delta_0+\delta_\pi}{4}~\eta(y) \, ,
\label{alpha}
\ee
where the function $\epsilon(y)$ is the periodic sign function and
$\eta(y)$ is the `staircase' function
\be
\eta(y)=2 q+1 \, ,
\;\;\;\;\;\;\;
q \pi R<y<(q+1) \pi R \, ,
\;\;\;\;\;\;\;
(q\in {\mathbf Z}) \, .
\label{staircase}
\ee
The local field redefinition that relates the two Lagrangians $\cL^O$
and $\cL^G$ is~\footnote{On the basis of the equivalence between
$\cL^O$ and $\cL^G$ shown here, we disagree with the statement of
ref.~\cite{mno} that `the Scherk-Schwarz mechanism is
equivalent to the mechanism of adding a mass term \ldots for sure not
when the mass terms are localized at the fixed points of the
orbifold'.}:
\be
\widetilde{\Psi}^G(y)={V^G}^\dagger (y) V^O(y)~ \widetilde{\Psi}^O(y)
\,.
\label{redefgo}
\ee
Notice that the periodic fields $\widetilde{\Psi}^G(y)$ are not smooth
but only piecewise smooth \cite{bfz1}. This can be checked either by
integrating the equations of motion for $\widetilde{\Psi}^G(y)$,
derived from $\cL^G$, in a small region around the fixed points
\cite{bfz1}, or by making use of the field redefinition in
Eq.~(\ref{redefgo}), recalling that $\widetilde{\Psi}^O(y)$ and
$V^O(y)$ are smooth while $V^G(y)$ is not.  We find that the fields
$\widetilde{\Psi}^G(y)$ have cusps and discontinuities described by:
\be
\left\{
\begin{array}{l}
\widetilde{\Psi}^G(2 q \pi R+\xi) = 
e^{\dd{\, i \, \delta_0 \sigma^2}}
\widetilde{\Psi}^G(2 q \pi R-\xi) 
\\
\widetilde{\Psi}^G[(2 q+1) \pi R+\xi] = 
e^{\dd{\, i \, \delta_\pi\sigma^2}} \,
\widetilde{\Psi}^G[(2 q+1) \pi R-\xi]
\end{array}
\right.
\, ,
\;\;\;\;\;
(0 < \xi \ll 1 \, ,
\, q \in {\mathbf Z} )
\, ,
\label{jumps}
\ee
where the `jumps' of the field variables are parametrized
by $\delta_{0,\pi}$.

Another simple but instructive example corresponds to a Lagrangian for
periodic fields of the form in Eq.~(\ref{result}), where now
\be
m_1(y)=m_2(y)=0 \, , 
\;\;\;\;\;
m_3(y) \ne 0 \, ,
\label{oddm}
\ee
and $m_3(y)$ is an otherwise arbitrary real, odd, periodic function of
$y$, as prescribed by Eqs.~(\ref{I})-(\ref{III}). Notice that, for
any such function, Eq.~(\ref{wloop}) gives always $\beta=0$, since
\be
\int_y^{y+2 \pi R} dy' m_3(y') = 0  \, .
\ee
In other words, real, periodic, odd mass profiles can be completely
removed by a field redefinition without introducing a non-trivial
twist. Such a field redefinition corresponds to:
\be
V(y) = \exp \left[ i \int_0^y dy' m_3(y') \right] \, .
\ee
Some representative profiles for $m_3(y)$ are exhibited in
Fig.~\ref{m3y}.
\FIGURE{
\epsfig{figure=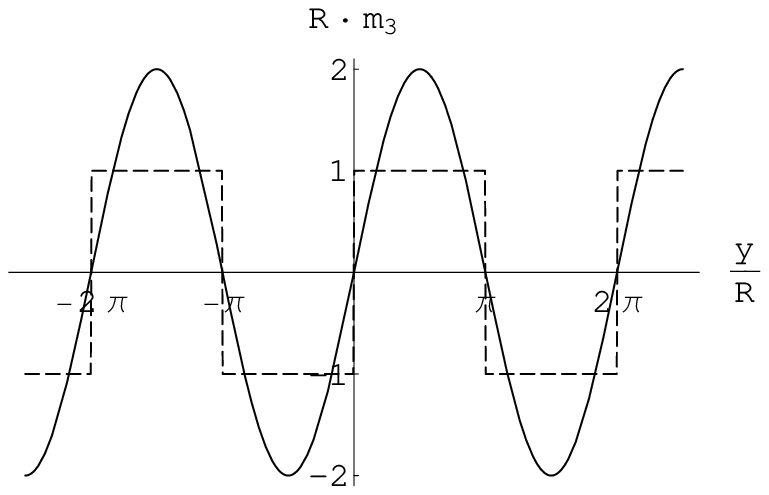,height=7.5cm}
\caption{Two representative and equivalent choices for $m_3(y)$: the
solid line corresponds to $m_3(y)=(2 \sin y)/R$, the dashed one to
$m_3(y)=\epsilon(y)/R$.}
\label{m3y}
}
Notice that no constant $m_3(y) \ne 0$ is allowed by
Eqs.~(\ref{I})-(\ref{III}), and also a $m_3(y) \ne 0$ completely
localized at $y = 2 q \pi R$ and/or $y = (2q + 1) \pi R$ is
forbidden. An allowed possibility is a piecewise constant $m_3(y)$,
for example:
\be
m_3(y) = \mu \, \epsilon(y) \, ,
\ee 
where $\epsilon(y)$ is the periodic sign function and $\mu$ a real
constant with the dimension of a mass.

Before concluding this section, it is appropriate to comment on the
relation between our $m_3(y)$ and other odd mass terms discussed in
related but different frameworks. Ref.~\cite{mno} discussed odd mass
terms that do not satisfy, in our notation, Eq.~(\ref{II}). Therefore,
the fact that those mass terms give a spectrum different from the one
of Eq.~(\ref{masses}) is not in contradiction with our results: we
have already stressed in Section~3 that the mass terms compatible with
the Scherk-Schwarz mechanism do not correspond to the most general set
of $y$-dependent mass terms allowed by 4D Lorentz invariance.

\section{Generalizations and discussion}

The results of the previous sections can be suitably generalized to 5D
theories with a general content of bosons and fermions, general
interactions, and general symmetries to be exploited for the
Scherk-Schwarz twist, as already discussed on some examples
\cite{bfz2,bf}. A particularly interesting case would be the effective
5D supergravity corresponding to M-theory \cite{hw}
compactified on a small Calabi-Yau manifold times a large orbifold
$S^1/Z_2$. In particular, it would be interesting to explore further
the intriguing analogies \cite{hor,aqgau} between non-perturbative
supersymmetry breaking via gaugino condensation at the orbifold fixed
points \cite{hor,bulk,mno1} and supersymmetry breaking via the
Scherk-Schwarz mechanism, in view of our generalized description
\cite{bfz1,bfz2} of the latter. The effective 5D supergravity of
M-theory is a quite complicated one, because of the presence of a
warped background and of different types of multiplets, including some
compactification moduli. Here we discuss only a toy example, namely
pure 5D Poincar\'e supergravity, showing that the general effective
theory of Scherk-Schwarz supersymmetry breaking can indeed encompass
\cite{bfz2,mq} the case of brane-induced supersymmetry breaking. To
conclude we briefly comment on the analogies and the differences with
the full-fledged M-theory.

In the on-shell Lagrangian of pure $N=1$, $D=5$ Poincar\'e
supergravity \cite{pure5d}, the 5D spinor of Eq.~(\ref{notation}) is
replaced by the 5D gravitino, in our notation:
\be
\Psi_M = \left(
\begin{array}{c}
\psi_{1M} \\
\psi_{2M}
\end{array}
\right) \, ,
\;\;\;\;\;
(M=m,5) \, .
\label{gravitino}
\ee
The bosonic superpartners of the gravitino are the f\"unfbein $E_M^{\;
A}$ and the graviphoton $B_M$, singlets under the global $SU(2)$
(R-)symmetry under which $\Psi_M$ transforms as in Eq.~(\ref{su2}),
with the same constant matrix $U$ for every value of the index $M$.
The $Z_2$ parity assignments to the fermionic fields are now $Z =
\widehat{\sigma}^3$ for $\Psi_m$, $Z = - \widehat{\sigma}^3$ for
$\Psi_5$, and can be consistently completed by assigning even parity
to $(E_m^{\; a}, E_5^{\; 5}, B_5)$ and odd parity to $(E_m^{\; 5},
E_5^{\; a}, B_m)$. We expand the theory around a flat background,
solution of the 5D equations of motion: 
\be
\langle E_M^{\; A} \rangle = \delta_M^A \, , 
\;\;\;\;\;
\langle \Psi_M \rangle = \langle B_M \rangle =0 \, .
\label{backg}
\ee

The implementation of the Scherk-Schwarz twist on $S^1/Z_2$ and the
derivation of the corresponding effective 5D theory for periodic
fields can be discussed in parallel with Sections~2 and 3, thus we do
not repeat all the details. Here we just comment on the new features
and on the resulting structure of 5D gravitino mass terms
(complementary details can be found in \cite{bfz2,mq}). The twisted
boundary conditions on the gravitino can be written as:
\be
\Psi_M (y+2 \pi R) = U_{\vec{\beta}} \, \Psi_M (y) \, ,
\;\;\;\;\;
(M=m,5) \, ,
\label{gtwist}
\ee
with $U_{\vec{\beta}}$ as in Eq.~(\ref{twisted}), and the
consistency conditions of Eqs.~(\ref{cc}) and (\ref{beta})
still apply. Similarly, the field redefinition bringing 
to the basis of periodic fields reads:
\be
\Psi_M (y)  =  V(y) \, \widetilde{\Psi}_M(y) \, ,
\;\;\;\;\;\;\;\;
(M=m,5) \, ,
\label{gredef}
\ee
where $V(y)$ satisfies as before the conditions of Eq.~(\ref{reqall}).

The only term of the Lagrangian involving derivatives of the gravitino
is its generalized kinetic term, which also involves the degrees of
freedom associated with the f\"unfbein.  Therefore, when moving to the
basis of periodic fields, not only mass terms but also some
interaction terms will be generated. Since we are interested here in
the structure of the gravitino mass terms, we set all the bosonic
fields to their background values of Eq.~(\ref{backg}) and focus on
the fermion bilinears. Up to a universal normalization factor, the
Lagrangian is
\be
\cL = - {1 \over 2} \epsilon^{mnpq} \ov{\Psi}_m^T \ov{\sigma}_n 
\partial_p \Psi_q + \left( - {i \over 2} \Psi_m^T \sigma^{mn} 
\widehat{\sigma}^2 \partial_y \Psi_n + i \Psi_5^T \sigma^{mn} 
\widehat{\sigma}^2 \partial_m \Psi_n
+ {\rm h.c.} \right) + \ldots \, .
\label{glag}
\ee
After moving to the basis of periodic fields via the field 
redefinition of Eq.~(\ref{gredef}), we get:
\bea 
\cL(\Psi_M,\partial\Psi_M) & = &
\cL(\widetilde{\Psi}_M,\partial\widetilde{\Psi}_M) 
+ \left\{
-\frac{i}{2}\left[m_1(y)+i \, m_2(y)\right] 
\widetilde{\psi}_{m1} \sigma^{mn} \widetilde{\psi}_{n1} 
\right. \nn \\ & + & \left.
\frac{i}{2}\left[m_1(y)-i \, m_2(y)\right]
\widetilde{\psi}_{2m} \sigma^{mn} \widetilde{\psi}_{2n} 
+ i \, m_3(y) \, \widetilde{\psi}_{1m} \sigma^{mn} 
\widetilde{\psi}_{2n} + {\rm h.c.} \right\} \, .
\label{gresult}
\eea
From here on the discussion can proceed as in the simpler case of the
previous sections. The only important difference is that gravitino
masses occur via the super-Higgs effect, with the goldstino components
provided by $\widetilde{\Psi}_5$. To discuss the spectrum in the case
of a non-trivial Scherk-Schwarz twist, $\beta \ne 0$, it is convenient
to go to the unitary gauge, where $\widetilde{\Psi}_5$ completely
disappears from the Lagrangian on the right-hand side of
Eq.~(\ref{gresult}).  We could now repeat the whole discussion of
Section~4. In particular, the case of Eqs.~(\ref{mgen}) and (\ref{mg})
corresponds to gravitino mass terms entirely localized at the orbifold
fixed points,
\be
\cL_{mass} (\widetilde{\Psi}_M,\partial\widetilde{\Psi}_M) 
= {1 \over 2} \left[ \delta_0 \, \delta(y) + \delta_\pi
\delta(y-\pi R) \right] \left( \widetilde{\psi}_{m1} 
\sigma^{mn} \widetilde{\psi}_{n1} + \widetilde{\psi}_{m2} 
\sigma^{mn} \widetilde{\psi}_{n2} + {\rm h.c.} \right) \, ,
\label{gloc}
\ee
where we can interpret the constants $\delta_0$ and $\delta_\pi$ as
the remnants of some localized brane dynamics, which may include
gaugino condensation. We remind the reader that \cite{bfz1,bfz2}
deriving the equations of motion and the mass spectrum in the presence
of the localized Lagrangian of Eq.~(\ref{gloc}) requires a
regularization~\footnote{Alternatively, one could use an equivalent
localized Lagrangian where only bilinears in the even fields do
appear, and to which one can apply the naive variational principle
without regularization. The precise meaning of such a Lagrangian is
discussed in \cite{bfz1,bfz2}, here we just stress that it cannot be
obtained from (\ref{gloc}) by means of an $SU(2)$ transformation.}.

We conclude with some comments on the possible extension of the
previous considerations to the case of gaugino condensation in
M-theory \cite{hor,bulk,mno1}. We recall that in such case localized
gravitino mass terms can be induced into the effective 5D supergravity
Lagrangian by the non-zero VEV of $G_{ABCD}$, the four-form of
eleven-dimensional supergravity. The VEV related with the gaugino
condensate, because of a perfect square structure that appears in the
Lagrangian, is $\langle G_{11abc} \rangle$, where $a,b,c = 1,2,3$ are
holomorphic indices associated with the six-dimensional Calabi-Yau
manifold. $\langle G_{11abc} \rangle$ is even under the $Z_2$ parity,
and generates gravitino mass terms of the type of $m_2(y)$ in
Eq.~(\ref{gresult}). The other possible VEV, $\langle G_{a \ov{a} b
\ov{b}} \rangle$, is odd under the $Z_2$ parity, and generates
gravitino mass terms of the form of $m_3(y)$ in
Eq.~(\ref{gresult}). However, now $m_3(y)$ is imaginary rather than
real as dictated by Eq.~(\ref{II}). As discussed in \cite{mno}, such a
term may give a spectrum different from the Scherk-Schwarz one.
Naively, this would suggest that, in the presence of a non-vanishing
$\langle G_{a\ov{a}b\ov{b}} \rangle$, gaugino condensation in M-theory
cannot be reinterpreted as a generalized Scherk-Schwarz mechanism.
However, there are other effects that play a role. For example, the
odd mass terms generated by $\langle G_{a\ov{a}b\ov{b}} \rangle$ only
appear in the intermediate steps of the derivation of the effective 5D
theory. In the final form of the resulting 5D supergravity Lagrangian,
as written for example in \cite{mno1}, the contribution of those mass
terms cancels, after an integration by parts, against a contribution
originated by the non trivial $y$-dependence of some moduli fields. We
conclude that the possible equivalence of gaugino condensation in
M-theory with a generalized Scherk-Schwarz mechanism is still an open
issue, whose complete clarification requires further work.

\bigskip

\acknowledgments
C.B. and F.F. thank the CERN Theory Division for its hospitality
during part of this project. F.Z. thanks the Physics Department of the
University of Padua for its hospitality during part of this project,
and INFN, Sezione di Padova, for partial travel support.  This work
was partially supported by the European Programmes HPRN-CT-2000-00149
(Collider Physics) and HPRN-CT-2000-00148 (Across the Energy
Frontier).

\bigskip

\bigskip

\appendix

\section{Appendix}
We collect here some useful formulae and results concerning
path-ordered products, and show how they can be used to prove
Eq.~(\ref{converse}) and Eq.~(\ref{IV}). First, we distinguish the two
inequivalent definitions of path-ordering, introducing the symbols:
\bea 
P_> [m(y_1) m(y_2)] & \equiv & 
m(y_1) m(y_2) \Theta(y_1 - y_2) + 
m(y_2) m(y_1) \Theta(y_2 - y_1) \, , 
\nn \\ 
P_< [m(y_1) m(y_2)] & \equiv & 
m(y_1) m(y_2) \Theta(y_2 - y_1) + 
m(y_2) m(y_1) \Theta(y_1 - y_2) \, ,
\label{Pdef}
\eea
where $m(y)$ is a $y$-dependent matrix and
\be
\Theta(y) = \left\{
\begin{array}{lcc}
1 & {\rm for} & y>0  \\
0 & {\rm for} & y<0
\end{array}
\right.
\label{theta}
\ee
is the Heaviside step function. From the above definitions, and
assuming $y_1 < y_2 < y_3$, the following properties follow:
\be
P_> \left[\exp\left({\dd i \, 
\int_{y_1}^{y_2} dy' m(y')}\right)\right] 
\, \cdot \,
P_<\left[\exp\left({\dd - i \, 
\int_{y_1}^{y_2} dy' m(y')}\right)\right] 
= {\mathbf 1} \, ,
\label{propa}
\ee
\be
P_> \left[\exp\left({\dd  i \!\! 
\int_{y_1}^{y_3} \!\! dy' m(y')}
\right)\right]
=
P_> \left[\exp\left({\dd  i \!\! 
\int_{y_2}^{y_3} \!\! dy' m(y')}
\right)\right]
\, \cdot \,
P_>\left[\exp\left({\dd  i \!\! 
\int_{y_1}^{y_2} \!\! dy' m(y')}
\right)\right]  
\, ,
\label{propb}
\ee
\be
P_< \left[\exp\left({\dd i \!\!
\int_{y_1}^{y_3} \!\! dy' m(y')}
\right)\right]
=
P_< \left[\exp\left({\dd i \!\! 
\int_{y_1}^{y_2} \!\! dy' m(y')}
\right)\right]
 \, \cdot \,
P_<\left[\exp\left({\dd i \!\! 
\int_{y_2}^{y_3} \!\! dy' m(y')}\right)\right]  
\, .
\label{propc}
\ee
If the $y$-dependent matrix $V(y)$ satisfies the differential
equation:
\be
\partial_y \, V(y) = i \, V(y) \, m(y) \, ,
\label{dv}
\ee
then it is immediate to prove that
\be
V(y) = V(y_0) \, P \left[\exp\left({\dd i \, 
\int_{y_0}^y dy' m(y')}\right)\right] \, ,
\;\;\;\;\;
P = \left\{
\begin{array}{ccc}
P_< & {\rm for} & y_0 < y \\
P_> & {\rm for} & y_0 > y
\end{array}
\right.
\, .
\label{Pv}
\ee
Correspondingly, if $m(y)$ is hermitian, then $V^\dagger(y)$ obeys the
equation
\be
\partial_y \, V^\dagger(y) = - i \, m(y) \,  V^\dagger(y) \, ,
\label{dvd}
\ee
which is solved by
\be
V^\dagger(y) = P \left[\exp \left( {\dd - i \, 
\int_{y_0}^y dy' m(y')} \right) \right] V^\dagger(y_0) 
\, ,
\;\;\;\;\;
P = \left\{
\begin{array}{ccc}
P_> & {\rm for} & y_0 < y \\
P_< & {\rm for} & y_0 > y
\end{array}
\right. \, .
\label{Pvd}
\ee
Showing that Eq.~(\ref{mcform}) implies Eq.~(\ref{converse}) is now a
simple application of Eqs.~(\ref{dv}) and (\ref{Pv}). To show instead
that Eqs.~(\ref{req1}) and (\ref{mcform}) imply Eq.~(\ref{IV}), it is
sufficient to solve Eq.~(\ref{req1}) for $U_{\vec{\beta}}$,
\be
U_{\vec{\beta}} = V(y + 2 \pi R) V^\dagger (y) \, ,
\ee
and to insert the explicit form of the solutions of
Eq.~(\ref{mcform}), namely Eqs.~(\ref{Pv}) and (\ref{Pvd}).  It is
also easy to show that the second member of Eq.~(\ref{IV}) is indeed
$y$-independent.
\newpage
\end{document}